\documentclass[onecolumn,showpacs,preprintnumbers,amsmath,amssymb,11pt]{revtex4}

\usepackage{epsf}
\usepackage{graphicx}  
\usepackage{dcolumn}   
\usepackage{bm}        

\newcommand{\be}{\begin{equation}}
\newcommand{\en}{\end{equation}}
\newcommand{\bea}{\begin{eqnarray}}
\newcommand{\ena}{\end{eqnarray}}
\begin{document}

\preprint{}

\title{Thermodynamics of the apparent horizon in massive cosmology}

 \author{Hui Li \footnote{Electronic address: lihui@ytu.edu.cn} }
 \affiliation{
 Department of Physics, Yantai University, 30 Qingquan Road, Yantai 264005, Shandong Province, P.R.China }

\author{Yi Zhang\footnote{Electronic address: zhangyia@cqupt.edu.cn}}
\affiliation{College of Mathematics and Physics, Chongqing University of
Posts and Telecommunications, Chongqing 400065, P.R.China}

\begin{abstract}
 Applying Clausius relation with energy-supply defined by the unified first law of thermodynamics formalism to the apparent horizon of a massive cosmological model proposed lately, the corrected entropic formula of the apparent horizon is obtained with the help of the modified Friedmann equations. This entropy-area relation, together with the identified internal energy, verifies the first law of thermodynamics for the apparent horizon with a volume change term for consistency. On the other hand, by means of the corrected entropy-area formula and the Clausius relation $\delta Q=T dS$, the modified Friedmann equations governing the dynamical evolution of the universe are reproduced with the known energy density and pressure of massive graviton. The integration constant is found to correspond to a cosmological term which could be absorbed into the energy density of matter. Having established the correspondence of massive cosmology with the unified first law of thermodynamics on the apparent horizon, the validity of the generalized second law of thermodynamics is also discussed by assuming the thermal equilibrium between the apparent horizon and the matter field bounded by the apparent horizon. It is found that, in the limit $H_c\rightarrow 0$ which recovers the Minkowski reference metric solution in the flat case, the generalized second law of thermodynamics holds if $\alpha_3+4\alpha_4<0$. Apart from that, even for the simplest model of dRGT massive cosmology with $\alpha_3=\alpha_4=0$, the generalized second law of thermodynamics could be violated.

\end{abstract}


\maketitle

\section{Introduction}

SNIa observations support a present accelerating universe\cite{acce}. With regard to general relativity(GR), a hypothetic dark energy component is necessary to meet the remarkable observations\cite{synthesis}. Cosmological constant is a simplest resolution in the framework of classical field theory; however, the surprisingly small value of the cosmological constant seems unnatural in light of quantum gravity, which is believed to take over the UV region of quantum fluctuations, remove the singularity problem and unify general relativity and quantum field theory at short distance. That means an infrared peculiarity to some extent is entangled with the UV divergence and the IR region should also be modified. Most of dark energy models have reasonable motivations and observational expectations; in the meantime, due to that cosmological constant problem\cite{Weinberg} and moreover the so-called cosmological coincidence problem, they only acquire limited success and are still far from satisfactory.  A second approach to understand the acceleration phenomenon relies on the modified gravity theories, such as theories of extra dimensions such as DGP models\cite{DGP} and massive gravity. Different from theories of extra dimensions where gravitons acquire mass through dimensional reduction to four dimensions, a tiny mass is endowed to the graviton simply by hand\cite{massive}. Interestingly enough, this deformation of general relativity can effectively give rise to a small cosmological constant term within, for instance, the simplest bimetric models of massive gravity\cite{bimetric}. It turns out that the graviton mass not only reproduces a cosmological term, but at the same time can manifests itself as other types of matter content with different equations of state\cite{cham}. In the linear model of massive gravity with Fierz-Pauli mass, the longitudinal graviton maintains a finite coupling to the trace of the source stress tensor even in the massless limit. This incurs the problem of vDVZ discontinuity\cite{vDVZ} which means the Fierz-Pauli model can not reduce to GR in the massless limit $m\rightarrow 0$ and therefore directly contradicts experiments on the solar system. By way of the Vainshtein mechanism\cite{Vainshtein} in the classical framework, the neglected non-linearity may be strong and the nature of non-linear instability helps to restore continuity with GR below the Vainshtein radius. The Lagrangian for the helicity-0 component generically contains nonlinear terms with more than two time derivatives; the latter give rise to the sixth degree of freedom on local backgrounds\cite{BDghost}. The presence of the Boulware-Deser (BD) ghost notoriously hinders us from constructing a healthy theory of Lorentz invariant massive gravity which recovers GR. Recently, de Rham, Gabadadze and Tolley (dRGT) have successfully constructed a non-linear model\cite{dRGT} of massive gravity which is ghost-free in the decoupling limit to all orders and furthermore at the complete non-linear level\cite{Hassan}. Therefore, dRGT gravity is well under investigations theoretically and observationally\cite{dRGTMin} as well.

To inspect a gravitational theory thermodynamical analysis has becoming a powerful tool. As pivotal events, blackhole thermodynamics\cite{bh} and recent AdS/CFT correspondence\cite{AdS} show explicit significance and strongly suggest the deep connection between gravity and thermodynamics.
A recent landmark of the identification of gravity theories and thermodynamics is the seminal work of Jacobson where the inverse problem of reproducing gravity theories from thermodynamical systems was seriously dealt with and successfully realized\cite{Jacobson}. By assuming the Clausius relation $\delta Q=T dS$ holds for all local Rindler causal horizons through each space-time point, Einstein field equations are deduced with the well-known entropy formula $S=A/(4G)$. The variation of heat flow $\delta Q$ is measured by an accelerated observer just inside the horizon and correspondingly $T$ denotes its Unruh temperature. Although the formulas were deduced in the null directions, it is suggested that the results may also be applied to all other directions in the tangential of the space-time. More recently, Eling and his collaborators discussed corresponding thermodynamical implications of $f(R)$ theories by means of similar method\cite{Eling}. To reproduce the correct equations of motion of $f(R)$ gravity, an entropic generating term should be added to the Clausius relation $\delta Q=TdS$ as well as the substitution of $S=\alpha f'(R)A$ to the entropy formula $S=A/(4G)$. It infers that $f(R)$ gravity is a non-equilibrium thermodynamics in essence. (See \cite{GongPRL} for a different viewpoint.) Along with this direction, various gravity theories have been checked and it is found that scalar-tensor theory of gravity also corresponds to non-equilibrium thermodynamics and an appropriate entropy production term is needed to derive the dynamic equations of motion space-times\cite{ScalarT}.

This theoretical complication entails examining the correspondence for different contexts besides for different gravity theories. For specific space-times in various gravitational theories, different strategies have been developed in the past few years. In the case of Einstein-Hilbert gravity, Einstein equations for a spherically symmetric space-time can be interpreted as the thermodynamic identity $dE=TdS-PdV$\cite{Padmana} with $S$ and $E$ being the entropy and energy derived by other approaches. What's more, the field equations for Lanczos-Lovelock action in a spherically symmetric space-time can also be expressed as the above form. As the modified terms could emerge in quantum pictures, it is remarkable to find thermodynamics can profile gravity beyond the classical level in this way. Another progress more relevant to our present work is the method of Hayward for dynamical blackholes\cite{Hayward,Hayward2}. In dealing with thermodynamics of a dynamical black hole in 4-dimensional Einstein theory, the associated trapping horizon is introduced for spherically symmetric space-times. In this formalism, Einstein field equations can be recast into the so-called unified first law and the first law of thermodynamics for the dynamical black hole is thus obtained by projecting the unified first law along a vector tangent to the trapping horizon\cite{CaoBW}. The change of local Rindler horizon to a topologically different trapping horizon which is globally geometric seems crucial for the thermodynamical reformulation of non-stationary space-times in various gravity theories\cite{Cai1}.

Our universe is also a non-stationary gravitational system which should be cautiously handled while carrying out thermodynamical analysis. In cosmological settings, the homogeneous and isotropic Friedmann-Robertson-Walker (FRW) metric is often assumed and the expanding 3-space is characterized by the cosmic scale factor which evolves with time. At a first glance, it appears that the FRW universe as one kind of dynamical spherically symmetric space-times can be easily dealt with by the method of unified first law. The subtlety occurs when we notice that, in the FRW universe, the (out) trapping horizon is absent. But fortunately, an inner trapping horizon still exists in cosmology. In the context of FRW metric, this horizon coincides with the apparent horizon and therefore, the apparent horizon is a natural choice in the foundation of thermodynamics. It stimulates a series of work on the foundation and discussion of the associated gravitational thermodynamics on this apparent horizon. In cosmology, apart from the apparent horizon, there exist many other special surfaces which are the Hubble horizon, the particle horizon, and the cosmological event horizon, etc.. And in certain cases they could coincide with one another. Therefore, it is interesting but difficult to know which one is appropriate for the formulation of the first law of thermodynamics. Due to a radical speculation that any surface in any space-time should have an entropy related to its area concurs with the entanglement entropy approach to dynamical blackholes\cite{Hayward}, it is believed that the question deserves deep investigations; however, it is not the point we discuss below and we will not step further on this in the present work. When focusing on the apparent horizon and the associated thermodynamics, in the setting of FRW universe, some authors investigated the relation between the first law and the Friedmann equations describing the dynamic evolution of the universe\cite{Frolov}. By applying the fundamental relation $\delta Q=T dS$ to the apparent horizon of the FRW universe, Cai and Kim derived the Friedmann equations\cite{Kim} with arbitrary spatial curvature. The Friedmann equations for the dynamical spherically symmetric space-times were also derived in the Gauss-Bonnet gravity and more general Lovelock gravity, where the actions of gravity theories are beyond Einstein theory with only a linear term of scalar curvature. When using the cosmological event horizon other than the apparent horizon in the calculation, the Friedmann equations describing the dynamics of the universe could only be obtained for the flat universe with $k=0$ FRW metric where the cosmological event horizon coincides with the apparent horizon (see also Ref.\cite{Wu}). In Ref.~\cite{AkbarWorkGL}, the derivation of the corresponding Friedmann equations by way of the first law of thermodynamics with a volume change term on the apparent horizon was also implemented for Einstein gravity, Gauss-Bonnet gravity and Lovelock gravity. For the scalar-tensor gravity and $f(R)$ gravity, the possibility to derive the corresponding Friedmann equations in those theories was investigated in \cite{ScalarT,AlbarWorkfR}.

In study of having established the first law of thermodynamics, it is usually propelled to test the validity of the second law of thermodynamics.
In the accelerating universe, for instance, the dominant energy condition may be violated and the second law of thermodynamics $\dot{S}_h>0$ does not hold any more. It is at this point a tentative version of the generalized second law of thermodynamics is proposed. The key idea is to assume that the thermal system bounded by the apparent horizon remains equilibrious and the temperature of the whole system is uniform; then the total entropy of the apparent horizon and the entropy of the matter fields inside the apparent horizon can be calculated with the well-founded settings of the first law of thermodynamics. The generalized second law of thermodynamics is often examined in this sense for the accelerating phase, viscous fluid and other exotic matter dominating\cite{Setare} universe and extended gravity theories such as Gauss-Bonnet gravity, Lovelock gravity\cite{Akbar3}, scalar-tensor theories\cite{Mazumder}, $f(R)$ theories\cite{Karami}, f(T) gravity\cite{Karami2}, Horava-Lifshitz cosmology\cite{Jamil}, modified f(R) Horava-Lifshitz gravity\cite{Chattopadhyay}, Gauss-Bonnet braneworld\cite{She}, warped DGP braneworld\cite{She2} and loop quantum cosmology\cite{Bandyopadhyay}.

Cosmological solutions of massive gravity with self-acceleration feature have been widely studied\cite{gong3} and it becomes appealing to explore the dark energy and dark matter problems in the framework of dRGT massive gravity. For the dRGT model, spatially open and flat de-Sitter solutions with an effective cosmological constant proportional to the graviton mass have been found. With certain evaluation of model parameters, the solutions with any spatial curvature also exist. Cosmological consequences have also been discussed in details; nevertheless, all those work assumes a Minkowski reference metric. Langlois and his collaborator proposed a slightly modified version of the original dRGT massive gravity in which the a priori arbitrary reference geometry is chosen to be de Sitter instead of Minkowski. Apart from the first two de-Sitter branches which were founded with the Minkowski reference metric, a third branch of self-accelerating solution has also been obtained\cite{Lang} and is subsequently studied in details in the literature\cite{gong4}. In this paper, we will examine the thermodynamical properties of such a cosmological model of dRGT massive gravity by the strategy elaborated in the work of Cai and Cao\cite{ScalarT}\cite{Cai-Cao}.

In this paper, dRGT massive cosmology with de-Sitter reference metric is introduced. Then, the Clausius relation with energy-supply defined by the unified first law of thermodynamics formalism is employed on the apparent horizon. With the help of the Friedmann equations, the corrected entropic formula of the apparent horizon is obtained. This entropy-area relation, together with the identified internal energy, verifies the first law of thermodynamics with a volume change term for consistency; secondly, by means of this corrected entropy-area formula and the Clausius relation $\delta Q=T dS$, where the temperature of the apparent horizon for energy crossing during the time interval $dt$ is $1/(2\pi \widetilde{r}_A)$ and the energy-supply of pure matter and the effective graviton energy density and pressure are expressed in terms of the Hubble parameter, the modified Friedmann equations governing the dynamical evolution of the universe are reproduced. The integration constant is found to correspond to a cosmological term which could be absorbed into the energy density of matter. Then, having established the correspondence of massive cosmology with the unified first law of thermodynamics on the apparent horizon, the validity of the generalized second law of thermodynamics is also discussed by assuming the thermal equilibrium between the apparent horizon and the matter field bounded by the apparent horizon. The temperature of the thermal system is therefore uniform and it could be appropriately handled to calculate the total entropy of the apparent horizon and the matter fields inside the apparent horizon. Finally we give the conclusion and discussions.

We start with the massive cosmology of dRGT gravity applying to homogeneous and isotropic space-time. The ghost free theory of massive gravity proposed by \cite{dRGT} is of the form
\begin{equation}
\label{action}
S=\frac{M_{pl}^2}{2}\int d^4x\sqrt{-g}(R+m_g^2\mathcal{U})+S_m,
\end{equation}
where $m_g$ is the mass of graviton, the nonlinear higher derivative terms for the massive graviton is
\begin{gather}
\label{massv}
\mathcal{U}=\mathcal{U}_2+\alpha_3\mathcal{U}_3+\alpha_4\mathcal{U}_4,\\
\label{massv2}
\mathcal{U}_2=[\mathcal{K}]^2-[\mathcal{K}^2],\\
\label{massv3}
\mathcal{U}_3=[\mathcal{K}]^3-3[\mathcal{K}][\mathcal{K}^2]+2[\mathcal{K}^3],\\
\label{massv4}
\mathcal{U}_4=[\mathcal{K}]^4-6[\mathcal{K}]^2[\mathcal{K}^2]+8[\mathcal{K}^3][\mathcal{K}]-6[\mathcal{K}^4],
\end{gather}
As dRGT construction points out, no higher order polynomial terms in $\mathcal{K}$ would exist and thus the most general Lagrangian density has only three free parameters, $m_g$, $\alpha_3$ and $\alpha_4$. The tensor $\mathcal{K}^\mu_\nu$ is
\begin{equation}
\label{tensor1}
\mathcal{K}^\mu_\nu=\delta^\mu_\nu-(\sqrt{\Sigma})^\mu_\nu,
\end{equation}
and $\Sigma_{\mu\nu}$ is defined by four St\"{u}ckelberg fields $\phi^a$ as
\begin{equation}
\label{stuckphi}
\Sigma_{\mu\nu}=\partial_\mu\phi^a\partial_\nu\phi^b\eta_{ab}.
\end{equation}
Usually the reference metric $\eta_{ab}$ is taken to be Minkowski. Recently, a different approach by choosing the priori arbitrary reference metric as de-Sitter instead of Minkowski has been proposed and in addition to the cosmological constant solutions, a new branch with much more sophisticated behavior has also been found. Specifically speaking, by varying the action with respect to the lapse function and scale factor, the Friedmann equations are obtained to be:
\begin{gather}
\label{frweq1}
H^2+\frac{k}{a^2}=\frac{1}{3 M_{pl}^2}(\rho_m+\rho_g),\\
\label{frweq2}
2\dot H+3H^2+\frac{k}{a^2}=-\frac{1}{M_{pl}^2}(p_m+p_g),
\end{gather}
where for the spatially flat $k=0$ case, the effective energy density $\rho_g$ and pressure $p_g$ for the massive graviton are\cite{gong4},
\begin{gather}
\label{effrho1}
\begin{split}
\rho_g=m_g^2M_{pl}^2\left[-6(1+2\alpha_3+2\alpha_4)+9(1+3\alpha_3+4\alpha_4)\frac{H}{H_c}\right.\\
\left.-3(1+6\alpha_3+12\alpha_4)\frac{H^2}{H_c^2}+3(\alpha_3+4\alpha_4)\frac{H^3}{H_c^3}\right],
\end{split}\\
\label{effp1}
\begin{split}
p_g=-\rho_g+m_g^2M_{pl}^2\frac{\dot H}{H^2}\frac{H}{H_c}\left[-3(1+3\alpha_3+4\alpha_4)+2(1+6\alpha_3+12\alpha_4)\frac{H}{H_c}\right.\\
\left.-3(\alpha_3+4\alpha_4)\frac{H^2}{H^2_c}\right].\\
\end{split}
\end{gather}
It is interesting to note that, when $H(z)=H_c$, $\rho_g$ equals zero, the energy density from massive graviton vanishes at this point.

\section{From Clausius relation and the modified Friedmann equations to the corrected entropy-area relation}

In this part, we will employ the Clausius relation with energy-supply defined by the unified first law of thermodynamics formalism on the apparent horizon ro obtain the corrected entropic formula of the apparent horizon. The modified Friedmann equations of dRGT cosmology with the spatially flat FRW metric will also be utilized. By regarding the introduction of the massive graviton as the deformation of Einstein gravity to dRGT massive gravity, it is appropriate to identify the contribution of massive graviton to be an effective energy-momentum part. Therefore, it can be reduced to the unified first law of thermodynamics of Einstein gravity and the energy-supply projecting along a vector $\xi$ tangent to the trapping horizon contains both the ordinary matter and the effective part from the massive graviton. After re-splitting the energy-supply term and presuming the heat flow of the Clausius relation to be the variation of heat flow $\delta Q$, the entropy of the apparent horizon can be obtained. As is implicitly meant in the unified first law of thermodynamics, the first law of thermodynamics for the apparent horizon still holds with a volume change term for consistency, and this point will also be checked with the resulting entropy-area relation and the identified internal energy.

Choosing $g_{\mu\nu}$ to be $n$-dimensional FRW metric:
\begin{eqnarray}
ds^2&=&g_{\mu\nu}dx^{\mu}dx^{\nu}=-dt^2+\frac{a(t)^2}{1-kr^2}dr^2+a(t)^2r^2d\Omega_{n-2}^2
\nonumber\\
&=&h_{ab}dx^a dx^b+\tilde{r}^2d\Omega_{n-2}^2\, ,
\end{eqnarray}
where $\tilde{r}=a(t)r$, $x^0=t, x^1=r$, $h_{ab}=diag(-1,a^2/(1-kr^2))$ with $k=-1, 0$ and $1$ for open, flat and closed spatial geometry respectively. The dynamical apparent horizon is defined to be the marginally trapped surface with vanishing expansion, and can be determined by the equality $h^{ab}\partial_a \tilde{r}\partial_b \tilde{r}=0$. Therefore, we can get the radius of the apparent horizon:
\begin{equation}
\widetilde{r}_A=\frac{1}{\sqrt{H^2+\frac{k}{a^2}}},
\end{equation}
where $H=\dot{a}/a$ is the Hubble parameter and the dot denotes the derivative with respect to cosmic time $t$. Differentiating the above equation with respect to the cosmic time $t$, it is obtained
 \begin{equation}
\dot{\widetilde{r}}_A=-H\widetilde{r}^3_A(\dot{H}-\frac{k}{a^2}).
\end{equation}
For the spatially flat $k=0$ case investigated bellow, the horizon radius and its evolution equations degenerate to be of the form
\begin{equation}
\widetilde{r}_A=\frac{1}{H}\label{rA}
\end{equation}
and
\begin{equation}
\dot{\widetilde{r}}_A=-\widetilde{r}^2_A\dot{H}.\label{rA2}
\end{equation}
That is, the apparent horizon coincides with the Hubble horizon in the spatially flat FRW case. 
Suppose that the energy-momentum tensor $T^{\mu\nu}$ of matter as well as the graviton has the form of a perfect fluid
$T^{\mu\nu}=(\rho+p)U^{\mu}U^{\nu}+p g^{\mu\nu}$, where $\rho$ and $p$ are the corresponding energy density and pressure respectively.
The energy conservation law is valid for the matter and graviton separately, and for the former it leads to the continuity equation
\begin{equation}
\dot{\rho}_m+3 H(\rho_m+p_m)=0,
\label{continuation}
\end{equation}
where the subscript denotes the quantities of matter in the universe throughout the paper by default.
Following Ref.\cite{Hayward}, the energy -supply vector $\Psi$ and the work density can be defined as
\begin{equation}
\Psi_a=T^b_a\partial_b \tilde{r}+W\partial_a\tilde{r},W=-\frac{1}{2}T^{a b}h_{a b}.
\end{equation}
where $T^{a b}$ is the projection of the $(3+1)$-dimensional energy-momentum tensor $T^{\mu \nu}$ in the normal direction of $2$-sphere of the FRW universe. For the present case, it is easy to find
\begin{equation}
\Psi=-\frac{1}{2}(\rho+p)H \widetilde{r} dt+ \frac{1}{2}(\rho+p)a dr,W=\frac{1}{2}(\rho-p)
\end{equation}

By means of the geometrical quantities of the area and volume of the $(n-2)$- sphere $A_{n-2}=\Omega_{n-2}\tilde r^{n-2}$ and $V_{n-2}=A_{n-2}
\tilde{r}/(n-1)$, the Misner-Sharp energy in $n$ dimensions inside
the apparent horizon of the FRW universe is written as~\cite{ScalarT}
\begin{equation}
E=\frac{1}{16\pi G_n}(n-2)\Omega_{n-2}\tilde{r}_A^{n-3}\, ,
\end{equation}
with $\tilde{r}_A$ the radius of the apparent horizon. Putting the $(00)$-component
of the equations of motion into the unified first law form, it reads
\begin{equation}
dE=A\Psi+WdV\, .
\end{equation}
Thus, the true first law of thermodynamics of the apparent horizon is obtained by projecting the above formula along a vector
$\xi=\partial_t-(1-2\epsilon)Hr\partial_r$ with $\epsilon=\dot{\tilde{r}}_A/(2H\tilde{r}_A$)\cite{CaoBW},
\begin{equation}
\langle dE, \xi\rangle=\frac{\kappa}{8\pi G}\langle dA,
\xi\rangle+\langle WdV, \xi\rangle\,.
\end{equation}
Note that $\kappa=-(1-\dot {\tilde r}_A/(2H\tilde r_A))/\tilde r_A$ is just the surface gravity of the apparent horizon.

We will derive an entropy expression associated with the apparent horizon of an FRW universe described by the modified Friedmann equations by using the method proposed in Ref.~\cite{CaoBW}. The energy-supply vector can be split into two parts:
\begin{equation}
\Psi=\Psi_m+\Psi_e
\end{equation}
with
\begin{equation}
\Psi_m=-\frac{1}{2}(\rho+p)H\tilde{r}dt+\frac{1}{2}(\rho+p)adr\,
\end{equation}
and
\begin{equation}
\Psi_e=-\frac{1}{2}(\rho_g+p_g)H\tilde{r}dt+\frac{1}{2}(\rho_g+p_g)adr.
\label{psie}
\end{equation}

The projection of the pure matter energy-supply $A\Psi_m$ on the apparent horizon supplies the heat flow $\delta Q$ in the Clausius
relation $\delta Q=T d S$. By using the unified first law of thermodynamics on the apparent horizon, there is
\begin{equation}
\delta Q\equiv \langle A\Psi_m,\xi \rangle= \frac{\kappa}{8\pi G}\langle dA,\xi \rangle-\langle A\Psi_e,\xi\rangle.
\label{Apsim}
\end{equation}
From Equations Eqns.~(\ref{effrho1}), ~(\ref{effp1}) and ~(\ref{psie}), we obtain
\begin{equation}
\langle A\Psi_m,\xi \rangle=-\frac{2\epsilon(1-\epsilon)}{G}-\frac{m_g^2}{G}\frac{\epsilon(1-\epsilon)\tilde{r}_A}{H_c}(-3\beta+2\gamma \frac{1}{H_c \widetilde{r}_A}-3\delta \frac{1}{H^2_c \widetilde{r}^2_A}).
\end{equation}
Assuming the temperature of the apparent horizon to be
\begin{equation}
T=\frac{\kappa}{2\pi},\label{T}
\end{equation}
the above equation can be recast into
\begin{equation}
\langle A\Psi_m,\xi \rangle=T\langle \frac{2\pi\widetilde{r}_A d\widetilde{r}_A}{G}+\frac{m^2_g}{G}\frac{\pi\widetilde{r}^2_A d\widetilde{r}_A}{H_c}(-3\beta+2\gamma \frac{1}{H_c \widetilde{r}_A}-3\delta \frac{1}{H^2_c \widetilde{r}^2_A}),\xi \rangle.\label{ds}
\end{equation}
Compared with the Clausius relation $\delta Q=T dS$, it is easy to accomplish the integration and obtain the corresponding entropy scaling which deviates from the usual $S=A/(4G)$; that is, we reach for the first time a corrected entropy-area relation in massive gravity:
 \begin{equation}
S=\frac{A}{4G}-\frac{m^2_g}{G}\frac{\beta}{H_c}\pi \tilde{r}^3_A+\frac{m^2_g}{G}\frac{\gamma}{H^2_c}\pi \tilde{r}^2_A -\frac{m^2_g}{G}\frac{3\delta}{H^3_c}\pi \widetilde{r}_A .\label{S}
\end{equation}
Note that we have introduced some new symbols of parameters for clarity and all of them are determined by the two free parameters of massive gravity:
\begin{equation}
\alpha=1+2\alpha_3+2\alpha_4,
\label{alpha}\emph{}
\end{equation}
\begin{equation}
\beta=1+3\alpha_3+4\alpha_4,
\label{beta}
\end{equation}
\begin{equation}
\gamma=1+6\alpha_3+12\alpha_4,
\label{gamma}
\end{equation}
\emph{}\begin{equation}
\delta=\alpha_3+4\alpha_4.
\label{delta}
\end{equation}
Therefore, the entropy of massive gravity does not observe the usual area law and the correction terms are all proportional to the square of the graviton mass. Once the mass of graviton approaches zero, the entropy-area relation reproduces the well-known result of Einstein gravity. Notice that, in all the terms of the entropy formula, the power exponents are positive integers which is clearly different from those of the Gauss-Bonnet gravity and the more general Lovelock gravity. For the latter cases, the blackhole entropy reads \cite{cai3}:
\begin{equation}
S=\frac{A}{4G} \Sigma_{i=1}^{m} \frac{i(n-1)}{(n-2i+1)} c_i r_{+}^{2-2i}
\end{equation}
where $A=n \Omega_n r_{+}^{n-1}$ is the horizon area of the black hole and $c_i$ are some coefficients.

Reasonably, by adopting the form of the total energy inside the apparent horizon to be $E_m=\rho_m V$, it is not difficult to verify the first law of thermodynamics for the apparent horizon with the entropy formula Eqn. (~\ref{S}),
\begin{equation}
dE_m=TdS+W_m dV
\end{equation}
Once again, we refer to the work density $W_m=(\rho_m-p_m)/2$ and volume of the apparent horizon $V=4/3 \pi \tilde{r}_A^3$.

\section{From the corrected entropy formula to modified Friedmann equations}

 In the above paragraph we have obtained the corrected entropy-area formula Eqn.~(\ref{S}). Let me refer to the assumptions to proceed: the heat flow $\delta Q$ is the energy-supply of pure matter projecting on the vector $\xi$ tangent to the apparent horizon and should be looked on as the amount of energy crossing the apparent horizon during the time interval $dt$; the temperature of the apparent horizon for energy crossing during the same interval $dt$ is $1/(2\pi \widetilde{r}_A)$. After reckoning on the substantial form of energy density and pressure of massive graviton in spatially flat dRGT FRW cosmology with de-Sitter reference metric, the modified Friedmann equations governing the dynamical evolution of the universe will be reproduced by way of the Clausius relation $\delta Q=T dS$.

Assuming the radius of apparent horizon $\tilde{r}_A$ constant, the amount of energy crossing the apparent horizon during the time internal $dt$ is approximately\cite{Kim}
\begin{equation}
\delta Q=-A\Psi_m=A(\rho_m+p_m)H\tilde{r}_A dt
\end{equation}
where $A=4\pi\tilde{r}^2_A$ is the area of the apparent horizon.

Moreover, suppose that the apparent horizon has an associated corrected entropy $S$ obtained above and temperature $T=1/(2\pi \widetilde{r}_A)$, the first law of thermodynamics of the above equation gives
\begin{equation}
A(\rho_m+p_m)H\widetilde{r}_A dt=\frac{1}{2\pi \widetilde{r}_A}(\frac{2\pi\widetilde{r}_A d\widetilde{r}_A}{G}+\frac{m^2_g}{G}\frac{\pi\widetilde{r}^2_A d\widetilde{r}_A}{H_c}(-3\beta+2\gamma \frac{1}{H_c \widetilde{r}_A}-3\delta \frac{1}{H^2_c \widetilde{r}^2_A})),\label{tds}
\end{equation}
With the help of Eqn.~(\ref{rA2}), it leads to
\begin{equation}
A(\rho_m+p_m)=\frac{-H \widetilde{r}_A^3 \dot{H}}{G}(1+m_g^2(\frac{-3\beta}{H_c}\frac{1}{2H}+\frac{\gamma}{H_c^2}-\frac{3\delta}{H_c^3}\frac{H}{2})),\label{tds2}
\end{equation}
As was stated above, the matter density $\rho_m$ satisfies the continuity equation individually,
\begin{equation}
\label{e29}
 \dot{\rho}_m+3 H(\rho_m+p_m)=0\, .\nonumber
\end{equation}
Therefore, the Clausius relation yields
\begin{equation}
\frac{8\pi G}{3}\dot{\rho}_m=2H \dot{H}(1+m_g^2(\frac{-3\beta}{H_c}\frac{1}{2H}+\frac{\gamma}{H_c^2}-\frac{3\delta}{H_c^3}\frac{H}{2})),\label{tds5}
\end{equation}
Integrating this equation yields
\begin{equation}
H^2=\frac{8\pi G}{3}\rho_m+m_g^2(\frac{3\beta}{H_c}H-\frac{\gamma}{H_c^2}H^2+\frac{\delta}{H_c^3}H^3))+C,\label{tds5}
\end{equation}
where $C$ is the integral constant. Compared with the Friedmann Eqn.~(\ref{frweq1}) of massive gravity the constant should be
\begin{equation}
C=2\alpha m_g^2
\end{equation}
Clearly the integration constant corresponds to a cosmological term and could be absorbed into the energy density of matter.
That fulfils the derivation of the modified Friedmann equations from the Clausius relation with the corrected entropy-area formula in massive cosmology.

\section{the generalized second law of thermodynamics}

 Together with previous systematic research on identifying the gravitational field equations with the first law of thermodynamics on the apparent horizon in various space-times, the calculation presented above once again indicates that the universality of the connection between gravity and thermodynamics can be enlarged to the case of massive gravity. It is of great interest to take a further step on the exploration of other thermodynamical aspects such as the tentative formulation of the thermodynamical second law in the settlings of massive cosmology. Having established the correspondence of massive cosmology with the unified first law of thermodynamics on the apparent horizon, it is not hard to compute the derivative of the entropy of the apparent horizon with respect to cosmic time.
Recall the modified Friedmann equations of massive gravity
\begin{equation}
H^2=\frac{1}{3M_{pl}^2}(\rho_m+m_g^2M_{pl}^2(-6\alpha+9\beta\frac{H}{H_c}-3\gamma\frac{H^2}{H_c^2}+3\delta\frac{H^3}{H_c^3})),\label{fried1}
\end{equation}
\begin{equation}
2\dot{H}+3H^2=-\frac{1}{M_{pl}^2}(p_m+m_g^2M_{pl}^2(6\alpha-9\beta\frac{H}{H_c}+3\gamma\frac{H^2}{H_c^2}-3\delta\frac{H^3}{H_c^3}+\frac{\dot{H}}{H^2}\frac{H}{H_c}(-3\beta+2\gamma\frac{H}{H_c}-3\delta\frac{H^2}{H_c^2}))).\label{fried2}
\end{equation}
Combined with Eqns.(~\ref{rA}) and (~\ref{rA2}), it is found that
\begin{equation}
-\frac{2\dot{\widetilde{r}}_A}{\widetilde{r}^3_A}=\frac{1}{3M_{pl}^2}(\dot{\rho}_m+m_g^2M_{pl}^2(9\beta\frac{-\dot{\widetilde{r}}_A}{\widetilde{r}^2_A H_c}+3\gamma\frac{2H\dot{\widetilde{r}}_A}{H_c^2 \widetilde{r}^2_A}-3\delta\frac{3H^2 \dot{\widetilde{r}}_A}{H_c^3 \widetilde{r}^2_A})).\label{Fried13}
\end{equation}
With respect to the continuity equation (~\ref{continuation}) of the matter density $\rho_m$, Eqn. ~(\ref{Fried13}) gives
\begin{equation}
\dot{\widetilde{r}}_A=\frac{\widetilde{r}^3_A}{2M_{pl}^2} H(\rho_m+p_m)\frac{1}{1+m_g^2(\frac{-3\beta}{2H H_c}+\frac{\gamma}{H_c^2}-\frac{3\delta H}{2H_c^3})}.\label{fried15}
\end{equation}
On the other hand, the associated temperature on the apparent horizon can be expressed in the form
\begin{equation}
T_h=\frac{|\kappa|}{2\pi}=\frac{1}{2\pi \widetilde{r}_A}(1-\frac{\dot{\widetilde{r}}_A}{2H\widetilde{r}_A})
\end{equation}
where $\dot{\widetilde{r}}_A/(2H\widetilde{r}_A)<1$ to ensure the positivity of the temperature.
Recognizing the entropy $S_h$ of the apparent horizon to be deduced through the connection between gravity and the first law of thermodynamics, we know that
\begin{equation}
T_h \dot{S}_h=\frac{1}{2\pi \widetilde{r}_A}(1-\frac{\dot{\widetilde{r}}_A}{2H\widetilde{r}_A})(\frac{2\pi\widetilde{r}_A \dot{\widetilde{r}}_A}{G}+\frac{m^2_g}{G}\frac{\pi\widetilde{r}^2_A \dot{\widetilde{r}}_A}{H_c}(-3\beta+2\gamma \frac{1}{H_c \widetilde{r}_A}-3\delta \frac{1}{H^2_c \widetilde{r}^2_A})) ; \label{ThSh}
\end{equation}
therefore
\begin{equation}
T_h \dot{S}_h=\frac{1}{2G}\dot{\widetilde{r}}_A(2-\dot{\widetilde{r}}_A)(1+\frac{m_g^2}{2}\frac{1}{H_c H}(-3\beta+2\gamma\frac{H}{H_c}-3\delta\frac{H^2}{H_c^2})).
\end{equation}
The two Friedmann equations~(\ref{fried1}) and ~(\ref{fried2}) can be recast into the following form
\begin{equation}
\rho_m+p_m=-2M_{pl}^2\dot{H}-\frac{\dot{H}}{H H_c}(-3\beta+2\gamma\frac{H}{H_c}-3\delta\frac{H^2}{H_c^2})m_g^2M_{pl}^2.\label{Fried-ch1}
\end{equation}
As a result, Eqn. (~\ref{ThSh}) bcomes
\begin{equation}
T_h \dot{S}_h=A(\rho_m+p_m)(1-\frac{\dot{\widetilde{r}}_A}{2})
\end{equation}
The positivity of the apparent horizon temperature requires $\dot{\widetilde{r}}_A<2$ in the spatially flat FRW case, and then the result means, without exotic matter components violating weak energy condition, the apparent horizon entropy always increases with time and the second law of thermodynamics holds in the whole history of cosmic expansion. However, in the accelerating universe the dominant energy condition is violated and the second law of thermodynamics $\dot{S}_h>0$ does not hold any more. It is at this point a tentative version of the generalized second law of thermodynamics is proposed. The key idea is to assume that the thermal system bounded by the apparent horizon remains equilibrious so that the temperature of the system is uniform across the boundary and then to consider the total entropy of the apparent horizon and the matter fields inside the apparent horizon. This requires that the temperature $T_m$ of the energy inside the apparent horizon should be the same as that of the apparent horizon; that is, $T_m=T_h$ throughout the whole evolution of the universe. A possible difference of the two temperatures would measure the spontaneous heat flow between the horizon and the matter inside it, which will not be dealt with in the present work.

The entropy of matter fields inside the apparent horizon, $S_m$, can be obtained by the Gibbs equation
\begin{equation}
T_m dS_m=d(\rho_m V)+p_m dV,
\end{equation}
where $E=\rho_m V$ is its energy and $p$ is its pressure in the horizon and
\begin{equation}
T_h \dot{S}_h+T_m \dot{S}_m=T\dot{S}_h+V\dot{\rho}_m+(\rho_m+p_m)\dot{V}.\label{genetropy}
\end{equation}
With regard to Eqn. (~\ref{Fried-ch1}), we have
\begin{equation}
A(\rho_m+p_m)=8\pi M_{pl}^2 \dot{\widetilde{r}}_A-4\pi \dot{\widetilde{r}}_A \frac{\widetilde{r}_A}{H_c}(-3\beta+2\gamma\frac{H}{H_c}-3\delta\frac{H^2}{H_c^2})m_g^2 M_{pl}^2,
\end{equation}
and the last two terms of the right hand side of Eqn.(~\ref{genetropy}) read
\begin{equation}
T_m \dot{S}_m=8\pi M_{pl}^2 \dot{\widetilde{r}}_A(\dot{\widetilde{r}}_A-1)(1-\frac{m_g^2}{2}\frac{1}{H_c H}(3\beta-2\gamma\frac{H}{H_c}+3\delta\frac{H^2}{H_c^2})).
\end{equation}
Therefore,
\begin{equation}
T_h \dot{S}_h+T_m \dot{S}_m=\frac{1}{2G}\dot{\widetilde{r}}^2_A(1-\frac{m_g^2}{2}\frac{1}{H_c H}(3\beta-2\gamma\frac{H}{H_c}+3\delta\frac{H^2}{H_c^2}))
\end{equation}
or
\begin{equation}
T_h \dot{S}_h+T_m \dot{S}_m=\frac{1}{2}A(\rho_m+p_m)\dot{\widetilde{r}}_A.
\end{equation}
Considering Eqn.~(\ref{fried15}), the evolution of the total entropy can be obtained:
\begin{equation}
T_h \dot{S}_h+T_m \dot{S}_m=2\pi G A \tilde{r}_A^2(\rho_m+p_m)^2\frac{1}{1+m_g^2(\frac{-3\beta}{2H H_c}+\frac{\gamma}{H_c^2}-\frac{3\delta H}{2H_c^3})},\label{wholeentropy}
\end{equation}

It is not hard to find that the generalized second law of thermodynamics with this setting does not always hold and its validity clearly depends on the signature of the denominator in Eqn.~(\ref{wholeentropy}) or all the free parameters $\alpha_3$, $\alpha_4$,$m_g$ and $H_c$. When the deforming parameter $m_g$ is vanishing, the present model naturally degenerates to the FRW cosmology in general relativity and the generalized second law of thermodynamics clearly holds. The last free parameter $H_c$ is an additional parameter which does not appear in the original dRGT massive cosmology. The de-Sitter reference metric brings about such a mass scale and in the limit $H_c\rightarrow 0$, one recovers the Minkowski reference metric solution in the flat case. In this limit, $\delta=\alpha_3+4\alpha_4$ should be negative for the generalized second law of thermodynamics to be valid. As for other cases, and even for the minimal massive cosmological model with $\alpha_3=\alpha_4=0$, the interplay of $m_g$ and other parameter(s) makes the situation complicated and the Higuchi bound may deeply involve in the discussion since there exists an absolute minimum for the mass of a spin-2 field set by such a bound in de Sitter space-time\cite{Higuchi}. It seems that, at present no such version of the generalized second law of thermodynamics in massive cosmology could be verified.

\section{Conclusion and Discussion}

Jacobson found that the connection between gravity and thermodynamics can be materialized by identifying Einstein field equation with the Clausius relation $\delta Q=TdS$. By assuming the space-time to be spherically symmetric, the thermodynamical relation which holds pointwise can be transferred to be associated with a globally geometric horizon. While applying this program to cosmological settings, the cosmological apparent horizon is employed as well as the unified first law of thermodynamics which primarily aims at the description of dynamical blackholes. These two theoretical elements have been incorporated into a systematic formulation which is elaborated in the work of Cai et.al.. In this paper, by means of that strategy, starting from the modified Friedmann equations in dRGT massive cosmology with de-Sitter reference metric, an entropy expression associated with the apparent horizon of an FRW universe is obtained by use of the unified first law projecting on the dynamical horizon and splitting the energy-supply term into the pure matter part and the effective energy-supply part. The form of the corrected entropy-area formula is clearly different from that of general relativity and more general Lovelock gravity. With this apparent horizon entropy-area formula, the first law of thermodynamics $dE=TdS+W_m dV$ is naturally satisfied in terms of the identified total energy $E$ and the work term in the unified first law of thermodynamics.

On the other hand, applying the Clausius relation $\delta Q=TdS$ to apparent horizon of a spatially flat FRW universe, and assuming that the apparent horizon has the temperature of $T=1/(2\pi \widetilde{r}_A)$, the observation that the pure matter energy-supply $A\Psi_m$ (after projecting along the apparent horizon) gives the heat flow $\delta Q$ in the Clausius relation directly leads to the modified Friedmann equations governing the dynamical evolution of the universe. The integration constant as well as other terms in the right hand side of the derived modified Friedmann equation reflects the fact that our present cosmological model of dRGT massive gravity exhibits a richer dynamical behavior. The effective gravitational fluid generated by the graviton mass not only contains a cosmological constant, but manifests itself as other types of matter content with different equations of state.

On the footing of the first law of thermodynamics which is verified, a further step is also taken for the comprehensive understanding of the thermodynamical properties of dRGT massive cosmology.  As is well-known to us, the second law of thermodynamics is not always satisfied for different fluids in various gravitational theories or in the accelerating universe. Together with the matter fields' entropy inside the apparent horizon, the generalized second law of thermodynamics was proven to hold in Gauss-Bonnet braneworld and in warped DGP braneworld and so on. By inspecting the evolution of the apparent horizon entropy deduced through the connection between gravity and the first law of thermodynamics, for massive gravity, we follow the strategy of the generalized second law and as a rudimentary calculation, we adopt their hypothesis that the thermal system remains equilibrious between the apparent horizon and the matter field inside the horizon. We found that the total entropy can decrease with time and this version of the generalized second law of thermodynamics seems invalid in some parameter space. To extract appropriate parameter evaluation scope deeply involves the complicated interplay of all the four free parameters in massive cosmology\cite{Higuchi} and no definite results exists at present. Lately, it is found that all homogeneous and isotropic backgrounds, as well as most of known spherically-symmetric inhomogeneous solutions, have an intrinsic instability which is irrelevant to the BD ghost\cite{instability}. Whether the violation of the generalized second law of thermodynamics  should be attributed to the incompleteness of the massive gravity theory or the absence of some new principles is thoroughly unclear and may be worth further investigations.




 {\bf Acknowledgments.}
 H. Li is supported by National Foundation of China under grant Nos. 11205131 and 10747155. Y. Zhang is supported by the Ministry of Science and Technology of  the National Natural Science Foundation of China key project under grant Nos. 11175270, 11005164, 11073005 and 10935013, CQ CSTC under grant No. 2010BB0408.

\end{document}